\documentclass[10pt,twocolumn,article]{IEEEtran} 
\usepackage{times,comment}
\usepackage{amsbsy}
\usepackage{latexsym} 
\usepackage{amssymb}
\usepackage{enumitem}
\usepackage{mathtools,xparse}
\usepackage{amsmath,amsfonts,graphicx,epsfig,amsthm,mathtools,diffcoeff}
 \usepackage{times,latexsym,bm,color}
\usepackage[ansinew]{inputenc}
\usepackage{algpseudocode, algorithm}
\usepackage{setspace}
\usepackage{textcomp}
\usepackage{parskip}

\allowdisplaybreaks
\begin{document}
\title{\huge Study of Linear Precoding and Power Allocation for Large Multiple-Antenna Systems with Coarsely Quantized Signals}
\author{Silvio F. B. Pinto and Rodrigo C. de Lamare \vspace{-2.00em} 
\thanks{The authors are with the Center for Telecommunications Studies (CETUC), Pontifical Catholic University of Rio de Janeiro, RJ, Brazil. R. C. de Lamare is also with the Department of Electronics, University of York, UK. Emails: silviof@cetuc.puc-rio.br, delamare@cetuc.puc-rio.br}}
\maketitle 
\begin{abstract}
This work studies coarse quantization-aware BD  (${\scriptstyle\mathrm{CQA-BD}}$) and coarse quantization-aware RBD (${\scriptstyle\mathrm{CQA-RBD}}$) precoding algorithms for large-scale MU-MIMO systems with coarsely quantized signals and proposes the coarse-quantization most advantageous allocation strategy (${\scriptstyle\mathrm{CQA-MAAS}}$) power allocation algorithm for linearly-precoded MU-MIMO systems. An analysis of the sum-rate along with studies of computational complexity is also carried out. Finally, comparisons between existing precoding and its power allocated version are followed by conclusions.
\end{abstract}

\section{Introduction}
Despite the evolution in signal processing techniques with 1-bit quantization \cite{Landau, Mezghani} for reducing power consumption in the large number of DACs used in massive MIMO systems, the achievable sum rates remain relatively low, which makes higher resolution quantizers with $b\geq 2$ bits attractive for the design of linear precoders and receivers. In these circumstances, Bussgang's theorem \cite{Bussgang} allows us express Gaussian precoded signals that have been quantized as a linear function of the quantized input and a distortion term which has no correlation with the input \cite{Sven1,Sven2,Sven3}. This approach enables the computation of the sum-rates of Gaussian signals \cite{Rowe}.

In this context, block diagonalization ${\scriptstyle\mathrm{\left(BD\right)}}$-type precoding methods \cite{Spencer1, Stankovic, Zu_CL, Zu, Sung,Wence} yield linear transmit approaches for multiuser MIMO (MU-MIMO) systems based on singular value decompositions (SVD), which provide excellent achievable sum-rates in the case of significant levels of multi-user interference and multiple-antenna users. ${\scriptstyle\mathrm{BD}}$ precoding is motivated by its enhanced sum-rate performance as compared to standard linear zero forcing (ZF) and minimum mean-square error (MMSE) precoders and its suitability for use with power allocation due to the available power loading matrix with the singular values that avoids an extra SVD. However, ${\scriptstyle\mathrm{BD}}$ has not been thoroughly investigated with coarsely quantized signals so far. Furthermore, existing linear ZF and MMSE precoding techniques that employ 1-bit quantization in massive MU-MIMO systems often present relatively poor performance and significant losses relative to full-resolution precoders. Additionally, precoding techniques in MU-MIMO systems can greatly benefit from power allocation strategies such as waterfilling. Specifically, power allocation can greatly enhance the sum-rate and error rate performance by employing higher power levels for channels with larger gains and lower power levels for poor channels. Previous works have considered iterative waterfilling techniques \cite{Yu2004}, practical algorihms \cite{Palomar} and specific strategies for ${\scriptstyle\mathrm{BD }}$ precoders \cite{Khan2014} even though there has been no power allocation strategy that takes into account coarse quantization so far, which could enhance the performance of precoders with low-resolution signals.

In this work, we investigate coarse quantization-aware BD  (${\scriptstyle\mathrm{CQA-BD}}$) and coarse quantization-aware RBD (${\scriptstyle\mathrm{CQA-RBD}}$) precoding algorithms for large-scale MU-MIMO systems with coarsely quantized signals and present the coarse-quantization most advantageous allocation strategy (${\scriptstyle\mathrm{CQA-MAAS}}$) power allocation algorithm for linearly-precoded MU-MIMO systems \cite{Pinto2}. An analysis of the sum-rate along with studies of computational complexity is also carried out. Numerical results illustrate the performance of the analyzed precoding and power allocation algorithms. 

This paper is organized as follows. Section II introduces the system model. Section III describes the ${\scriptstyle\mathrm{CQA-BD}}$ and ${\scriptstyle\mathrm{CQA-RBD}}$ precoding algorithms. Section IV presents the ${\scriptstyle\mathrm{CQA-MAAS}}$ power allocation. Section V presents the numerical results, whereas Section VI gives the conclusions.

\section{System Model}
\label{sysmodel}
\begin{figure}[ht] 
	\centering 
	\includegraphics[width=7.4cm, height=3.5cm]{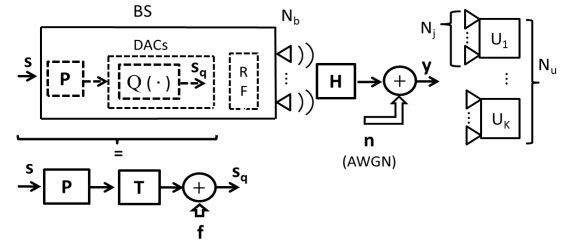} 
	\vspace{-1.5em}
	\caption{Outline of a quantized massive MU-MIMO broadcast system. Upper diagram: some simplified parts of BS. Lower diagram: Bussgang's theorem applied to the detached part of interest.}
	\label{sysmodel_CQA_BD}
\end{figure}
Let us consider the broadcast channel (BC) of a MU-MIMO system with a BS containing $N_{b}$ antennas, which sends radio frequency (RF) signals to users equipped with a total of $N_{u}=\sum_{j=1}^{K}\: N_{j}$ receive antennas, where $N_{j}\geq 1$ denotes the number of receive antennas of the $j$th user $U_{j}$, $ j=1,\ldots,K $, as outlined in Fig. \ref{sysmodel_CQA_BD}. 
%

The input-output relation of the BC can be modelled as
\begin{equation}
\label{downlink_channel_model}
\mathbf{y}= \mathbf{H}\;\mathbf{s}_{q}\:+\:\mathbf{n},
\end{equation}
where $\mathbf{y} \in \mathbb{C}^{N_{u}} $ contains the signals received by all users and $\mathbf{H} \in \mathbb{C}^{N_{u} \times N_{b}} $ stands for the matrix which models the assumed broadcast channel that is assumed known to the BS. The entries of $\mathbf{H}$ are considered independent circularly-symmetrical complex Gaussian random variables $ \left[ \mathbf{H}\right]_{u,b}\in \mathbb{CN} \left( 0,1\right) $, $u=1,\cdots, N_{u}$ and $b=1,\cdots, N_{b}$. The noise vector $\mathbf{n} \in\mathbb{C}^{N_{u}} $  is characterized by its i.i.d.  circularly-symmetric complex Gaussian entries $n_{u}\in \mathbb{CN} \left( 0,N_{0}\right)$. The noise variance is known at the BS and so is the sampling rate of  DACs at BS and ADCs at user equipments. 
Following the lower part of Fig.\ref{sysmodel_CQA_BD}, the quantization $\mathrm{Q} \left( \cdot\right)$ of a precoded symbol vector $\mathbf{Ps}$, where $\mathbf{P}\in \mathbb{C}^{Nb\times Nu}$ is a precoding matrix and $\mathbf{s} \sim \mathbb{CN}\left( \mathbf{0}_{Nu\times1},\mathbf{I}_{Nu} \right) $ is the symbol vector, can be expressed by the quantized vector given by
\begin{equation}
\label{bus_theo1}
\mathbf{s}_{q}=\mathrm{Q}\left(\mathbf{Ps}\right)= \mathbf{TPs+f},
\end{equation} 
where the  $\mathbf{f}$ and the symbol $\mathbf{s}$ vectors are uncorrelated. For approximations of achievable sum-rates involving $N_{b}$ and $N_{u}$  sufficiently large, it can be approximated as Gaussian noise, i.e., $\mathbf{f} \in\mathbb{C}^{N_{b}} $.

\section{Proposed CQA-BD and CQA-RBD Precoding Algorithms}

Both proposed  ${\scriptstyle\mathrm{CQA-BD}}$ and ${\scriptstyle\mathrm{CQA-RBD}}$ and their respective locally optimized variations ${\scriptstyle\mathrm{CQA-BD-MAAS}}$ and
${\scriptstyle\mathrm{CQA-RBD-MAAS}}$, respectively, are based on ${\scriptstyle\mathrm{BD}}$ and ${\scriptstyle\mathrm{RBD}}$ precoders $\mathbf{P}$ \cite{Spencer1,Sung, Stankovic,Zu,Zu_CL,Pinto2} given by
\begin{equation}
\label{conj_precod_matrix}
\mathbf{P}=\left[\mathbf{P}_{1} \mathbf{P}_{2}\cdots\mathbf{P}_{K}\right]\:\in \mathbb{C}^{N_{b} \times N_{u}}
\end{equation}
 where the precoding matrices $ \mathbf{P}_{j}$ for the \textit{j}th users  can be expressed as the product
\begin{equation}
 \label{precod_as_product}
 \mathbf{P}_{j}=\mathbf{P}_{j}^{c}\mathbf{P}_{j}^{d} 
 \end{equation}
 where $ \mathbf{P}_{j}^{c} \in \mathbb{C}^{N_{b} \times L_{j}}$ and $ \mathbf{P}_{j}^{d} \in \mathbb{C}^{L_{j} \times N_{j}}$. The parameter $L_{j}$ depends on which precoding algorithm is chosen, namely, the ${\scriptstyle\mathrm{ CQA-BD}}$ or ${\scriptstyle\mathrm{CQA-RBD}}$ techniques.
 
We can express the combined channel matrix $ \mathbf{H} $ and the resulting precoding matrix $ \mathbf{P} $ as follows:
\begin{equation}
\label{comb_ch_matrix}
 \mathbf{H}=\left[\mathbf{H}_{1}^{T} \mathbf{H}_{2}^{T}\cdots\mathbf{H}_{K}^{T}\right]^{T}\:\in \mathbb{C}^{N_{u} \times N_{b}}
\end{equation}
\begin{equation}
\label{conj_precod_matrix}
\mathbf{P}=\left[\mathbf{P}_{1} \mathbf{P}_{2}\cdots\mathbf{P}_{K}\right]\:\in \mathbb{C}^{N_{b} \times N_{u}}
\end{equation}
where $ \mathbf{H}_{j} \in \mathbb{C}^{N_{j} \times N_{b}} $ is the channel matrix of the $j$th user. The matrix $\mathbf{P}_{j} \in \mathbb{C}^{N_{b} \times N_{j}} $ represents the precoding matrix of the $j$th user.
\subsection{CQA-BD Precoder}

In the proposed ${\scriptstyle\mathrm{CQA-BD}}$ precoding algorithm, the first factor  in \eqref{precod_as_product} is given by 
\begin{equation}
\label{precod_mat_first_BD}
\mathbf{P}_{j}^{c\left(CQA-BD \right) }=\overline{\mathbf {W}}_{j}^{\left( 0\right) }
\end{equation}
where $ \overline{\mathbf{W}}_{j}^{\left(0 \right) } $ is obtained by the SVD \cite{Zu} of  \eqref{comb_ch_matrix}, in which the channel matrix of the $j$th  user has been removed, i.e.:
\begin{align}
\label{channel_matrix_user_exclusion}
\overline{\mathbf{H}}_{j}&=\left[\mathbf{H}_{1}^{T}\cdots \mathbf{H}_{j-1}^{T}\mathbf{H}_{j+1}^{T}\cdots\mathbf{H}_{K}^{T}\right]^{T}\:\in \mathbb{C}^{\overline{N}_{j} \times N_{b}}\nonumber\\&= \overline{\mathbf{U}}_{j}\overline{\mathbf{\Phi}}_{j}\overline{\mathbf{W}}_{j}^{H}=\overline{\mathbf{U}}_{j}\overline{\mathbf{\Phi}}_{j}\left[\overline{\mathbf{W}}_{j}^{\left( 1\right) }\overline{\mathbf{W}}_{j}^{\left(0 \right) } \right]^{H} 
\end{align}
where $\overline{N}_{j} =N_{u}-N_{j} $. 
The matrix $ \overline{\mathbf {W}}_{j}^{\left( 0\right) }
 \in \mathbb{C}^{{N}_{b} 
	\times \left( N_{b}-\overline{L}_{j}\right)}$, where $\overline{L}_{j}$ is the rank of $\overline{\mathbf{H}}_{j}$, uses the last $ N_{b}-\overline{L}_{j} $ singular vectors.

The second precoder of ${\scriptstyle\mathrm{CQA-BD}}$ in \eqref{precod_as_product} is obtained by SVD  of the effective channel matrix for the $j$th user $\mathbf{H}_{e_{j}}$ and employs a power loading matrix as follows:
\begin{align}
\label{second_factor_BD}
\mathbf{P}_{j}^{d\left(CQA-BD \right) )}=\mathbf{W}_{j}^{\left(1 \right)}\:{\left(\mathbf{\Omega}^{\left(CQA-BD \right)}_{j}\right)^\frac{1}{2}} 
\end{align}
where the power loading matrix $\mathbf{\Omega}^{\left(CQA-BD \right)}_{j}$ requires a power allocation algorithm and the matrix $\mathbf{W}_{j}^{\left(1 \right)}$ incorporates the first $\Lambda_{e}=rank\left( \mathbf{H}_{e_{j}}\right)$ singular vectors obtained by the decomposition of $ \mathbf{H}_{e_{j}} $, as follows: 
\begin{align}
\label{effect_channel_matrix}
\mathbf{H}_{e_{j}}&= \mathbf{H}_{j} \mathbf{P}_{j}^{c}\nonumber= \mathbf{U}_{j}\mathbf{\Phi}_{j}\mathbf{W}_{j}^{H}\\&=\mathbf{U}_{j}
\begin{bmatrix}
\mathbf{\Phi}_{j} & 0\\ 0 & 0
\end{bmatrix}
\begin{bmatrix}
\mathbf{W}_{j}^{\left(1 \right)}&\mathbf{W}_{j}^{\left(0 \right)}
\end{bmatrix}^{H}
\end{align}
\subsection{CQA-RBD Precoder}

In the case of the proposed ${\scriptstyle\mathrm{CQA-RBD}}$ precoding algorithm,  the first precoder in \eqref{precod_as_product} is given \cite{Stankovic,Zu} by
\begin{align}
\label{precod_mat_first_RBD}
\mathbf{P}_{j}^{c\left(CQA-RBD \right)}= \overline{\mathbf{W}}_{j} \left(\overline{ \mathbf{\Phi}}_{j}^{T} \overline{ \mathbf{\Phi}_{j}} + {\chi}\;\mathbf{I}_{N_{b}}
\right)^{-1/2} 
\end{align}
where $ \chi = \frac{N_{u}\sigma_{n}^{2}}{P} $ is the regularization factor required by the ${\scriptstyle\mathrm{CQA-RBD}}$ algorithm  and $ P $ is the average transmit power. 

The second precoder of ${\scriptstyle\mathrm{CQA-RBD}}$ in \eqref{precod_as_product} is obtained by SVD  of the effective channel matrix for the $j$th user $\mathbf{H}_{e_{j}}$ and power loading, respectively as follows:
\begin{align}
\label{second_factor_RBD}
\mathbf{P}_{j}^{d\left(CQA-RBD \right) )} = \mathbf{W}_{j}\:{\left(\mathbf{\Omega}^{\left(CQA-RBD \right)}_{j}\right)^\frac{1}{2}}
\end{align}
where the matrix $\mathbf{W}_{j}^{\left(1 \right)}$ incorporates the early $ \Lambda_{e}=rank\left( \mathbf{H}_{e_{j}}\right)$ singular vectors obtained by the decomposition of $ \mathbf{H}_{e_{j}} $, as follows: 
\begin{align}
\label{effect_channel_matrix_1}
\mathbf{H}_{e_{j}}&= \mathbf{H}_{j} \mathbf{P}_{j}^{c}\nonumber= \mathbf{U}_{j}\mathbf{\Phi}_{j}\mathbf{W}_{j}^{H}\\&=\mathbf{U}_{j}
\begin{bmatrix}
\mathbf{\Phi}_{j} & 0\\ 0 & 0
\end{bmatrix}
\begin{bmatrix}
\mathbf{W}_{j}^{\left(1 \right)}&\mathbf{W}_{j}^{\left(0 \right)}
\end{bmatrix}^{H}
\end{align}
The power loading matrix per user {$\mathbf{\Omega}_j^{\left(CQA-RBD \right)}$} can be obtained by a procedure like water filling (WF) \cite{Paulraj} power allocation and will be initialized with equal power allocation. 

The quantized vector \eqref{bus_theo1} combined with  the asumptions that  $\left( \mathit{N}_{b},\: \mathit{N}_{u}\right)$ is sufficiently large and that the quantization  error  from  DACs is Gaussian  lead \cite{Pinto2} to the following transmit processing matrix: 
\begin{equation}
\label{approx_diag_mat0}
\mathbf{T}_{n,n}= \delta\:\mathbf{I}_{Nb\times Nb},
\end{equation} 
where the scalar factor is described by
\begin{equation}
\delta= \alpha \gamma\sqrt{\frac{N_{b}}{\pi P}} \sum_{l=1}^{J-1}\exp\left(-\frac{N_{b}\gamma^{2}}{P}\left( 1-\frac{J}{2} \right)^{2} \right)\, 
\label{entries_diag_mat_dist} 
\end{equation}
which concentrates all process of quantization on the scalar $\delta$ in \eqref{approx_diag_mat0} and is used to compute the sum-rates at the receiver. The losses of achievable sum-rates for a fixed SNR due to the coarse quantization and a fixed realization of the channel are fully characterized by $\mathbf{T}_{n,n}$. The steps needed to compute ${\scriptstyle\mathrm{CQA-BD}}$ and ${\scriptstyle\mathrm{CQA-RBD}}$ are summarized in Algorithm \ref{algorithm:CQA_BD_RBD}.

\begin{algorithm}[htb!]
	\scriptsize
	\caption{Proposed CQA-BD and CQA-RBD  precoders } \label{algorithm:CQA_BD_RBD}	
	\begin{algorithmic}[1]
		\Require  
	$\begin{aligned} 
		\alpha&= \left( 2\mathrm {N_{b}}\gamma^{2} \left(\left( \frac{J-1}{2}\right)^{2} \right.\right.\nonumber\\&\left.\left. -2\sum_{l=1}^{J-1}  \left( 1-\frac{J}{2} \right)  \Xi \left( \sqrt{2N_{b}\gamma^{2}}\left( 1-\frac{J}{2} \right)\right)\right)\right)^{-1/2} \eqref{normalization_factor}
		\end{aligned} $ 
		$\begin{aligned}
		\delta= \alpha \gamma\sqrt{\frac{N_{b}}{\pi P}} \sum_{l=1}^{J-1}\exp\left(-\frac{N_{b}\gamma^{2}}{P}\left( 1-\frac{J}{2} \right)^{2} \right)\eqref{entries_diag_mat_dist}
		\end{aligned} $ 
         $\begin{aligned}
         \mathbf{H}=\left[\mathbf{H}_{1}^{T} \mathbf{H}_{2}^{T}\cdots\mathbf{H}_{K}^{T}\right]^{T}\:\in \mathbb{C}^{N_{u} \times N_{b}} \:\eqref{comb_ch_matrix}
		\end{aligned} $ 
		\For{$\mathrm{j = 1\;}\colon \: K$}
		\State $\overline{\mathbf{H}}_{j}=\left[\mathbf{H}_{1}^{T}\cdots \mathbf{H}_{j-1}^{T}\mathbf{H}_{j+1}^{T}\cdots\mathbf{H}_{K}^{T}\right]^{T}\:\in \mathbb{C}^{\overline{N}_{j} \times N_{b}}$\eqref{channel_matrix_user_exclusion}
		\State $\overline{\mathbf{H}}_{j}=  \overline{\mathbf{U}}_{j}\overline{\mathbf{\Phi}}_{j}\overline{\mathbf{W}}_{j}^{H}=\overline{\mathbf{U}}_{j}\overline{\mathbf{\Phi}}_{j}\left[\overline{\mathbf{W}}_{j}^{\left( 1\right) }\overline{\mathbf{W}}_{j}^{\left(0 \right) } \right]^{H}$ \eqref{channel_matrix_user_exclusion}
		\State $\mathbf{P}_{j}^{c\left(CQA-BD \right) }=\overline{\mathbf {W}}_{j}^{\left( 0\right) }$\eqref{precod_mat_first_BD}
		\State $\mathbf{P}_{j}^{c\left(RBD \right) }=\overline{\mathbf{W}}_{j}\left( \overline{\mathbf{\Phi}}_{j}^{T} \overline{\mathbf{\Phi}}_{j} + {\chi}\: \mathbf{I}_{N_{b}} \right)^{-1/2} $\eqref{precod_mat_first_RBD}
		\State $ 
		\mathbf{H}_{e_{j}}= \mathbf{H}_{j} \mathbf{P}_{j}^{c}= \mathbf{U}_{j}\mathbf{\Phi}_{j}\mathbf{W}_{j}^{H}=\mathbf{U}_{j} \begin{bmatrix}
		\mathbf{\Phi}_{j} & 0\\ 0 & 0
		\end{bmatrix}
		\begin{bmatrix}
		\mathbf{W}_{j}^{\left(1 \right)}&\mathbf{W}_{j}^{\left(0 \right)}
		\end{bmatrix}^{H}\nonumber $\eqref{effect_channel_matrix}
		
		\State {$\left(\mathbf{\Omega}^{\left(CQA-BD,CQA-RBD \right)}_{j}\right)^\frac{1}{2} $}  \text{by classical WF \cite{Paulraj} or variations} 
		
		\State $
		\mathbf{P}_{j}^{d\left(CQA-BD \right) )}=\mathbf{W}_{j}^{\left(1 \right)}\:{\left(\mathbf{\Omega}^{\left(BD \right)}_{j}\right)^\frac{1}{2}} $ \eqref{second_factor_BD}
		\State $
		\mathbf{P}_{j}^{d\left(CQA-RBD \right) )}=\mathbf{W}_{j}\:{\left(\mathbf{\Omega}^{\left(CQA-RBD \right)}_{j}\right)^\frac{1}{2}} $ \eqref{second_factor_RBD}
		\State $\mathbf{P}_{j}=\mathbf{P}_{j}^{c}\mathbf{P}_{j}^{d}$\:\eqref{precod_as_product}
		
		\EndFor
		\State $ \mathbf{P}=\left[\mathbf{P}_{1} \mathbf{P}_{2}\cdots\mathbf{P}_{K}\right]^{T}\:\in \mathbb{C}^{N_{b} \times N_{u}}  $ \eqref{conj_precod_matrix}
		
	\end{algorithmic}
\end{algorithm}

Extensions to other precoders and/or beamforming strategies \cite{wence,1bitcpm,dynovs,sint,sint2,rmmseprec,bapls,rmmsecf,mbthp,rmbthp,rsbd,rsthp,wljio,rdrcb,locsme,okspme,lrcc,armo,baplnc,jpaba} are possible. Moreover, detection and parameter estimation strategies can also be considered for future work \cite{jidf,jio,rrser,spa,mfsic,mbdf,bfidd,1bitidd,1bitadap,1bitce,listmtc}.

\section{Proposed CQA-MAAS Power Allocation}
The achievable rate \cite{Pinto} in bits per channel use at which information can be sent with arbitrarily low probability of error can be bounded by the mutual information of a Gaussian channel \cite{Cover,Telatar,Pinto} as follows: 
\begin{align}
\label{achievable_sum_rate_BD_RBD}
\mathit{C}\leq &\mathit{I}\left(\mathbf{s},\mathbf{y}\:\right) = \log_{2}\left\lbrace\det\left[ \mathbf{I}_{Nu} +  \frac{\mathit{SNR}}{\mathit{N}_{u}}\mathbf{\left( HP\right) } \mathbf{\left( HP\right) }^{H}\right.\right.\nonumber\\&\left.\left. \left(\left(1-\delta^{2} \right)\frac{\mathit{SNR}}{\mathit{N}_{u}}\mathbf{\left( HP\right) } \mathbf{\left( HP\right) }^{H} +\mathbf{I}_{Nu}              \right)^{-1}                 \right]  \right\rbrace 
\end{align}
where the  scalar factor $\delta $ concentrates the quantization impact.

The proposed CQA-MAAS power allocation involves appproximations of Neumann's (matrix), EVD and SVD in addition to other properties \cite{Seber, Harville},   which allow us to formulate \cite{Pinto2} it as a conditioned  maximization process of the achievable sum rate \eqref{achievable_sum_rate_BD_RBD}, as follows:
\begin{align}
\label{form_maximization_CQA_BD}
 \mathit{C}&\approx  \max_{\Phi_{j}} \sum_{j=1}^{K} \log_{2}
 \bigg|
  \mathbf{I}_{N_j}+\frac{\delta^{2}}{\mathrm{N}_{0}}\:\mathbf{\Phi}_{j}^{2}\:\mathbf{\Omega }_{j}  -\frac{\delta^{2}\left(1-\delta^{2} \right)}{\mathrm{N}_{0}^{2}}\:\mathbf{\Phi}_{j}^{4}\:\mathbf{\Omega }_{j}^{2}
  \bigg|\nonumber\\
  & = \max_{\Phi_{j}} \sum_{j=1}^{K} \mathrm{Tr}\left[\mathrm{\log}_{2}
 \bigg|
  \mathbf{I}_{N_j}+\frac{\delta^{2}}{\mathrm{N}_{0}}\:\mathbf{\Phi}_{j}^{2}\:\mathbf{\Omega }_{j}  \right.\nonumber\\& \left.- \frac{\delta^{2}\left(1-\delta^{2} \right)}{\mathrm{N}_{0}^{2}}\:\mathbf{\Phi}_{j}^{4}\:\mathbf{\Omega }_{j}^{2}
  \bigg| \right] \nonumber\\
  &\qquad \mathrm{s.t.}  \sum_{j=1}^{K} \mathrm{Tr} \left(\mathbf{\Omega}_{j}\right)\leq\:\mathit{P}_{total}
  \end{align}
  where $\mathbf{\Phi}_{j}$ and $\mathbf{\Omega}_{j}$ are estimated by the SVD of the non-interfering block channels \eqref{effect_channel_matrix} for ${\scriptstyle\mathrm{CQA-BD-MAAS}}$ and  \eqref{effect_channel_matrix_1} for ${\scriptstyle\mathrm{CQA-RBD-MAAS}}$.

The optimized precoding matrix for ${\scriptstyle\mathrm{BD }}$ algorithm $\mathbf{P}_{opt}$ makes use of a conditioned  maximization process of the achievable sum rate and involves appproximations of Neumann's (matrix) and  Mc Laurin's series   \cite{Seber, Harville}.  It is computed at each realization of the channel by incorporation of the power loading effects of  diagonal matrices $\Omega_{j}$, $ j=1,\ldots,K $, corresponding to the $N_{j}$ antennas of each $j$th user as follows:
\begin{align}
\label{optimal_precoding}
\mathbf{P}_{opt}^{BD}&=\left[\overline{\mathbf {W}}_{1}^{\left( 0\right) }\mathbf{W}_{1}^{\left(1 \right)}\:\left(\mathbf{\Omega}^{\left(BD \right)}_{1}\right)^\frac{1}{2} \cdots\overline{\mathbf {W}}_{K}^{\left( 0\right) }\mathbf{W}_{K}^{\left(1 \right)}\:\left(\mathbf{\Omega}^{\left(BD \right)}_{K}\right)^\frac{1}{2}\right]\nonumber\\
&=\left[\overline{\mathbf {W}}_{1}^{\left( 0\right) }\mathbf{W}_{1}^{\left(1 \right)} \cdots\overline{\mathbf {W}}_{K}^{\left( 0\right) }\mathbf{W}_{K}^{\left(1 \right)}\right]\:\left(\mathbf{\Omega}^{\left(BD \right)}\right)^{\frac{1}{2}}
\end{align}
where  $\mathbf{\Omega}^{\left(BD \right)}$  is a larger power diagonal matrix, where each of its $\mathrm{Nu}$ entries is  associated to its corresponding $\mathit{j}$th user, in ascending order, as follows:
  \begin{align}
   \label{}
   \mathbf{\Omega}^{\left(BD\right)}= \mathrm{diag}\{\mathbf{\Omega}_{1}, \cdots,\mathbf{\Omega}_{K} \},
   \end{align}
which can  be detailed as follows:
\begin{scriptsize}
\begin{align}
\label{Diag_power_matrix_exp}
 \mathbf{ \Omega}^{\left(BD\right)}= 
  \begin{bmatrix}
 \omega_{1} &  \cdots & 0 &\cdots\cdots & 0  &\cdots&  0\\
    \vdots&  \ddots & \vdots & & \vdots &\ddots &\vdots \\
    0 & \underbrace{\cdots}_{\mathbf{\Omega}_{1}} & \omega_{Nj}& \cdots\cdots &0 &\cdots  & 0 \\ \vdots & & \vdots &\ddots &\vdots & &\vdots\\\vdots & & \vdots& &\vdots & &\vdots\\0 &  \cdots & 0 &\cdots\cdots & \omega_{\left(\mathrm{N_{u}}-Nj\right)}  &\cdots&  0\\
    \vdots&  \ddots & \vdots & & \vdots &\ddots &\vdots \\
    0 & \cdots & 0 & \cdots\cdots &0 &\underbrace{\cdots}_{\mathbf{\Omega}_{K}}  & \omega_{\mathit{\left(N_{u}=K \times Nj\right)}}
 \end{bmatrix}
 \end{align}
 \end{scriptsize}
  The computation of  $\mathbf{\Omega}^{\left(BD \right)}$ is   based on  a locally optimized level of energy $\mu_{opt}$ \eqref{power_level_solution_implem}
    \begin{align}
   \label{power_level_solution_implem}
  \mu_{opt}&=\frac{\left(\mathit{N}_{u}-\mathit{p}+1\right)^{2}}{2\mathit{C}_{2}\mathrm{SNR}\sum_{m=1}^{\left(\mathit{N}_{u}-\mathit{p}+1\right)}\left[\phi^{2}\right]_{m}}\nonumber \\&\times \Bigg\{ 1 - \left[ 1 + \frac{4\mathit{C}_{2}}{\mathit{N}_{u}^{2}}\sum_{m=1}^{\left(\mathit{N}_{u}-\mathit{p}+1\right)}\left[\phi^{2}\right]_{m} \right.\nonumber\\& \times \left. \left( -\mathrm{SNR}\:+ \mathit{C}_{1} \sum_{m=1}^{\left(\mathit{N}_{u}-\mathit{p}+1\right)}\left[\frac{1}{\phi^{2}}\right]_{m}\right)                                     \right]^{\frac{1}{2}}\Bigg\},
   \end{align}
   where
   \begin{enumerate}[label=(\roman*)]
   \vspace{-2pt}
       \item \label{implem_1}$\mathit{N}_{u}$ stands for the number of receive antennas defined in Section \ref{sysmodel}.
       \item \label{implem_2}$\mathit{p} $ denotes an auxiliary parameter to be set to $1$.
       \item \label{implem_3}$\mathit{C}_{1}$ and $\mathit{C}_{2}$, which depend only on the distortion factor $\delta$, are given  by \eqref{c1} and \eqref{c2}.
       \begin{align}
   \label{c1}  
   \mathit{C}_{1}=\frac{\delta -\sqrt{4-3\delta^{2}}}
   {2\delta \left( 1-\delta^{2} \right)}
  \end{align}
  \begin{align}
   \label{c2} 
  \mathit{C}_{2}=\frac{\delta\left(1-\delta^{2} \right)}
   {\sqrt{4-3\delta^{2}}}
  \end{align}
       \item \label{implem_4} $\phi$ designates each of the $\mathrm{N}_{u}=\mathit{K \times N_{j}}$ singular values corresponding to each receive antenna. This can be better visualized in  the following diagonal matrix \eqref{Diag_sv_matrix}, in which the diagonal vector displays the required entries $\phi_{m}$ $\in$ $\phi_{1}, \cdots,\phi_{\mathit{N_{u}}}$  .

 \end{enumerate}
Employing the value of $\mu_{opt}$ provided by \eqref{power_level_solution_implem}, the power allocated to the $\mathit{m}$th $\in$ $\{1, \cdots,\mathrm{N_{u}}\} $ receive antenna can be computed by

   \begin{align}
  \label{single_power_expression_for_alg}
   \omega_{m} &= \mathit{C}_{1}\frac{\left(\mathit{N}_{u}-\mathit{p}+1\right)}{\mathrm{SNR}}\:\frac{1}{\phi^{2}_{m}} + \mu_{opt} \nonumber\\  &-\mu^{2}_{opt}\mathit{C}_{2}\frac{\mathrm{SNR}}{\left(\mathit{N}_{u}-\mathit{p}+1\right)}\:\phi^{2}_{m} 
  \end{align}
 where the parameters involved were defined in \ref{implem_1}, \ref{implem_2}, \ref{implem_3} and \ref{implem_4}. 
  Assuming that the power alloted to the receive antenna which is associated to the  minimum gain is negative, i.e., $\omega_{\mathit{N}_{u}-\mathit{p}+1}< 0$, it is rejected, and the algorithm must be executed with the parameter $\mathit{p}$ increased by unity. The 
  most advantageous allotment strategy is achieved at the time that  the power distributed among each receive antenna is non-negative according to the Khun-Tucker conditions \eqref{kt_condit}:
   \begin{eqnarray}
    \left.\left(\omega_{nn}\right)^{+}\right|_{nn=1,..., N_{u}}=\omega, &  \mathrm{if} \: \omega_{nn}\geq 0 \nonumber\\
        0, & \mathrm{if} \: \omega_{nn}< 0
  \label{kt_condit}
  \end{eqnarray}
The proposed  ${\scriptstyle\mathrm{CQA-MAAS}}$ power allocation can be summed up as in Algorithm \eqref{algorithm:Alg_power_allot}, as follows: %
\begin{algorithm}[htb!]
	 \scriptsize
		\caption{Proposed CQA-MAAS power allocation } \label{algorithm:Alg_power_allot}	
	{\begin{algorithmic}[1]
		\vspace{0.5em}
		\State $ \mathbf{Initialization}: \mathrm{Nu}, \mathrm{Nb}, \mathit{K} $\\
		\begin{scriptsize}
 \vspace{0.6em}
  \begin{align}
 \hspace{-40mm}
 \mathbf{\Phi}= 
 \begin{bmatrix}
 \phi_{1} &  \cdots & 0 \\
    \vdots&  \ddots & \vdots\\
    0&\cdots&\phi_{\mathit{N_{u}}} \nonumber
 \end{bmatrix}\eqref{Diag_sv_matrix}
\end{align}
\end{scriptsize}
		\State
		$\mathbf{Compute}$ the factors $\mathit{C}_{1}$ in \eqref{c1} and $\mathit{C}_{2}$ in \eqref{c2}. \vspace{0.4em}
		\State
		$\mathbf{Compute}$ the optimum energy level $\mu_{opt} $ in \eqref{power_level_solution_implem}.\vspace{0.5em}
		\State 
		$\mathbf{Compute}$ the power allocation to each sub-channel $\omega_{m} $ in \eqref{single_power_expression_for_alg}.\vspace{0.5em}
		\State 
		 $\mathbf{If}$ there are negative values, then find their minimum, i.e., their $\min \left(\omega_{N_{u}-p+1}<0\right) $ $\mathbf{and}$. 
		 \State 
		 $\mathbf{Refuse}$ this minimum negative value by assuming it is equal to zero in \eqref{kt_condit}, $\mathbf{and}$. \vspace{0.5em}
		 \State
		 $\mathbf{Perform}$ the algorithm with the parameter $\mathit{p}$ incremented by unity. \vspace{0.5em}
		 \State
		 CQA-MAAS achieves its goal when the power allocated among the receive antennas is non-negative \eqref{kt_condit}. \vspace{0.5em}
		 \State
		 $\mathbf{Compute}$ the power diagonal matrix \eqref{Diag_power_matrix_exp}, by relating its $N_{u}$ power entries to their corresponding receive antennas.
		\end{algorithmic}}
\end{algorithm} 
\section{Numerical results}
\begin{figure}[htb!]
	\centering 
	\includegraphics[width=8.2cm,height=6.4cm]{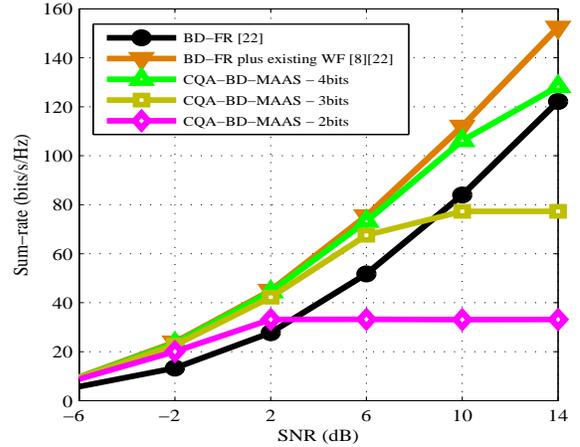}
	\vspace{-1em}
	\caption{Achievable rates for CQA-BD-MAAS, $2,3,4,$ quantization bits, via Bussgang theorem, Gaussian signals, compared to BD-FR and BD-FR-plus existing WF. MU-MIMO configuration: $ N_{b}=64 $  and $ N_{u}=16 \times 2 $. }
	\label{Base_2bits_blue_64_32_100_WSA_2021_2}
\end{figure}

We consider two scenarios with a MU-MIMO system using $ N_{b}=64 $  and $ N_{u}=16 \times 2 $ and$ N_{b}=64 $  and $ N_{u}=8 \times 2 $  that assume the conditions described in Section \ref{sysmodel}. Fig.\ref{Base_2bits_blue_64_32_100_WSA_2021_2} employs the firsf scenario in order to illustrate the achievable sum-rates for the proposed ${\scriptstyle\mathrm{CQA-BD-MAAS}}$, which results from coarse-quantization most advantageous allocation strategy (${\scriptstyle\mathrm{CQA-MAAS}}$) power allocation algorithm to block-diagonalization precoding for 2, 3 and 4-bit quantization. They are compared to BD full resolution ${\scriptstyle\mathrm{BD-FR}}$  and its existing variant  ${\scriptstyle\mathrm{BD-FR\: plus\: existing\: Waterfilling}}$ \cite{Spencer1}. Considering the influence of practical aspects, we have modeled an imperfect channel knowledge combined with a spatial correlation $\hat{\mathbf H}=\mathbf {H}\;\tilde{\mathbf{R}}\ ^{\frac{1}{2}} + \mathbf {E}$, where $\tilde{\mathbf{R}}$ represents the complex transmit correlation matrix \cite{Loyka,Zu} whose elements are
\begin{align}
\tilde{R}_{ij}=\left\{\begin{array}{ll} r^{j-i},& i\leq j\\ r_{ji}^*,& i>j \end{array} \right.,|r|\leq 1
\end{align}
where $\lvert\mathrm{r}\rvert <1$. It can be noticed that the absolute values of the entries $\lvert \tilde{\mathrm{R}}_{\left(i,j\right)} \rvert$ corresponding to the closest antennas are larger than the others. The error matrix  $\mathbf{E}$ is modeled \cite{Zu} as a complex Gaussian noise with i.i.d entries of zero mean and variance $\sigma_{e}^{2}$. In our next examples, we have employed large values of correlations between the neighboring antennas, i.e., $\lvert\mathrm{r}\rvert=0.72$ and $ 0.91 $, respectively. The variance  $\sigma_{e}^{2}$ of the  feedback error matrix  $\mathbf{E}$ has been set to $0.16$. 

\begin{figure}[htb!]
	\centering 
	\includegraphics[width=8.5cm,height=6.7cm]{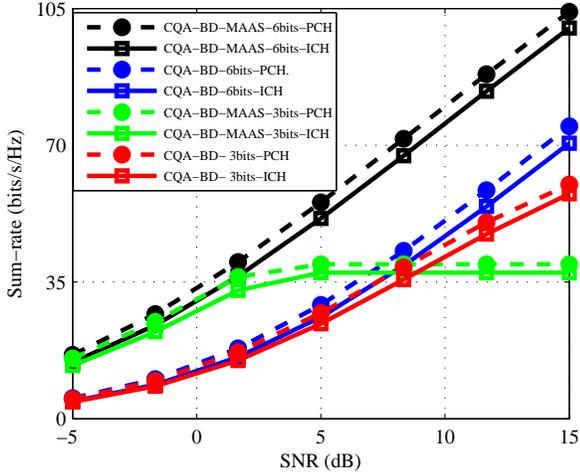}
	\caption{{Achievable rates for CQA-BD and  CQA-BD-MAAS for 6 and 3-bit quantization under perfect and imperfect channel knowledge (ICH). MU-MIMO configuration: $ N_{b}=64 $  and $ N_{u}=8 \times 2 $.
	 {$\lvert\mathrm{r}\rvert=0.72$  and $\sigma_{e}^{2}=0.16$.}}}
	\label{Channel_robustness_BD_3b_6b_v1}.
	\end{figure}

In Fig. \ref{Channel_robustness_BD_3b_6b_v1}, in which the scenario is composed with $ N_{b}=64 $  and $ N_{u}=8\times 2 $, we assess the performance of ${\scriptstyle\mathrm{CQA-BD}}$ and ${\scriptstyle\mathrm{CQA-BD-MAAS}}$ in the presence of imperfect channel knowledge and spatial correlation using $3$ and $6$ bits. The results show that the impact of imperfect channel knowledge is not significant in terms of performance degradation of the precoders. However, the performance degradation of ${\scriptstyle\mathrm{CQA-BD-MAAS}}$ can become significant for $3$ bits. 

 The number of \text{FLOPs} required by conventional ${\scriptstyle\mathrm{BD}}$ and ${\scriptstyle\mathrm{RBD}}$ algorithms are dominated by two \text{SVDs} \cite{Spencer1}. Since our system model is dedicated to broadcast channels, we can assume the widespread ratios $N_{b}\gg N_{u} \gg N_{j} $ and one of their resulting approximations $N_{u}- N_{j} \approx N_{u} $ to simplify the resulting expressions. Table \ref{table2} illustrates the computational cost required by the proposed ${\scriptstyle\mathrm{CQA}}$ and existing precoders.

\begin{table}[htp]\scriptsize
\caption{Computational complexity of  BD, RBD and their proposed coarsely quantized  variations based on Bussgang Theorem: CQA-BD and CQA-RBD}
\centering 
{\begin{tabular}{l l }
\hline\hline
Precoder & Computational cost (FLOPs) under $N_{b}\gg N_{u} \gg N_{j} $\\
\hline 
\\
BD  &  ${N_{b}^2}\left(32N_{j}+8\right) +N_{b}\left(32 N_{u}^{2} +72N_{j}^{2}\right)+ 64N_{u}^{2} $ \\ 
RBD  &  ${N_{b}^2}\left(32N_{j}+8\right) +N_{b}\left(32 N_{u}^{2} +72N_{j}^{2}\right)+ 64N_{u}^{2} $ \\
Proposed  & $ {N_{b}^2}\left(32N_{j}+8\right) +N_{b}\left(32 N_{u}^{2} +72N_{j}^{2}\right)+ 64N_{u}^{2} +C_{\delta}$ \\
CQA-BD &\\
Proposed  & $ {N_{b}^2}\left(32N_{j}+8\right) +N_{b}\left(32 N_{u}^{2} +72N_{j}^{2}\right)+ 64N_{u}^{2}+C_{\delta} $ \\
CQA-RBD &\\
\hline 
\end{tabular}}
\label{table2}
\end{table}
The extra cost $C_{\delta}$ required to convert  ${\scriptstyle\mathrm{BD}}$ and  ${\scriptstyle\mathrm{RBD}}$ into their corresponding  Bussgang-based precoders, which are listed in Table \ref{table2}, do not have significant impact on the total computational cost of their respective Bussgang-based algorithms. Due to their design, existing waterfilling and the proposed CQA-MAAS power allocation have a similar computational cost of $\mathcal{O}\left( N_{u}\right)$, which in practice does not result in significant additional cost to be imposed on ${\scriptstyle\mathrm{BD}}$ and ${\scriptstyle\mathrm{RBD}}$ to obtain their respective ${\scriptstyle\mathrm{CQA-BD-MAAS}}$ and ${\scriptstyle\mathrm{CQA-RBD-MAAS}}$ schemes. Table \eqref{table3}  summarizes these additional costs.

\begin{table}[htp]\small
\caption{Computational complexity of proposed CQA-MAAS and existing WF algorithms.}
\centering 
{\begin{tabular}{l c } 
\hline\hline 
 Technique & Computational cost (FLOPs) \\
\hline 
Waterfilling (WF) & $\mathcal{O}\left( N_{u}\right) $   \\ 
MAAS&  $\mathcal{O}\left( N_{u}\right) $   \\ 
\hline 
\end{tabular}}
\label{table3}
\end{table}

\section{Conclusion}
We have investigated ${\scriptstyle\mathrm{CQA-BD}}$ and ${\scriptstyle\mathrm{CQA-RBD}}$ precoding and developed the ${\scriptstyle\mathrm{CQA-RBD-MAAS}}$ power allocation algorithms for large-scale MIMO systems that employ coarse quantization using DACs with few bits.  ${\scriptstyle\mathrm{CQA-RBD-MAAS}}$ can obtain gains in sum-rate of up to $30$ \% over schemes without power allocation and comparable performance to full-resolution schemes with precoding and WF power allocation. The proposed algorithms can be used in massive MIMO systems and contribute to substantial reduction in power consumption.

\end{document}